\documentclass{article}

\PassOptionsToPackage{numbers, compress}{natbib}

\usepackage[final]{neurips_2020_ml4ps}

\usepackage[utf8]{inputenc} 
\usepackage[T1]{fontenc}    
\usepackage{hyperref}       
\usepackage{url}            
\usepackage{booktabs}       
\usepackage{amsfonts}       
\usepackage{nicefrac}       
\usepackage{microtype}      
\usepackage{amsmath}
\usepackage{subfig}
\usepackage{graphicx}
\graphicspath{{figures/}}

\title{Scalable, End-to-End, Deep-Learning-Based Data Reconstruction Chain for Particle Imaging Detectors}

\author{%
  Fran\c cois Drielsma\\
  SLAC National Accelerator Laboratory\\
  Menlo Park, CA 94025 \\
  \texttt{drielsma@stanford.edu} \\
  \And
  Kazuhiro Terao \\
  SLAC National Accelerator Laboratory \\
  Menlo Park, CA 94025 \\
  \texttt{kterao@stanford.edu} \\
  \And
  Laura Domin\'e \\
  Stanford University \\
  Stanford, CA 94305 \\ 
  \texttt{ldomine@stanford.edu} \\
  \And
  Dae Heun Koh \\ 
  Stanford University \\
  Stanford, CA 94305 \\
  \texttt{koh0207@stanford.edu} \\
}

\begin{document}

\maketitle

\begin{abstract}
    Recent inroads in Computer Vision (CV) and Machine Learning (ML) have motivated a new approach to the analysis of particle imaging detector data. Unlike previous efforts which tackled isolated CV tasks, this paper introduces an end-to-end, ML-based data reconstruction chain for Liquid Argon Time Projection Chambers (LArTPCs), the state-of-the-art in precision imaging at the intensity frontier of neutrino physics. The chain is a multi-task network cascade which combines voxel-level feature extraction using Sparse Convolutional Neural Networks and particle superstructure formation using Graph Neural Networks. Each algorithm incorporates physics-informed inductive biases, while their collective hierarchy is used to enforce a causal structure. The output is a comprehensive description of an event that may be used for high-level physics inference. The chain is end-to-end optimizable, eliminating the need for time-intensive manual software adjustments. It is also the first implementation to handle the unprecedented pile-up of dozens of high energy neutrino interactions, expected in the 3D-imaging LArTPC of the Deep Underground Neutrino Experiment. The chain is trained as a whole and its performance is assessed at each step using an open simulated data set.
\end{abstract}

\section{Introduction}
In recent years, the accelerator-based neutrino physics community in the United States has moved to employ Liquid Argon Time Projection Chambers (LArTPCs) as its central neutrino detection technology~\cite{rubbia_lartpcs, sbn, dune}. Charged particles that traverse the detector ionize the noble liquid. The electrons so-produced are drifted in a uniform electric field towards a readout plane. The location of electrons collected on the anode, combined with their arrival time, offers mm-scale resolution images of charged particle interactions, along with precise calorimetric information~\cite{icarus_cngs, larpix}.

\begin{figure*}[t]
    \centering
    \includegraphics[width=0.98\textwidth]{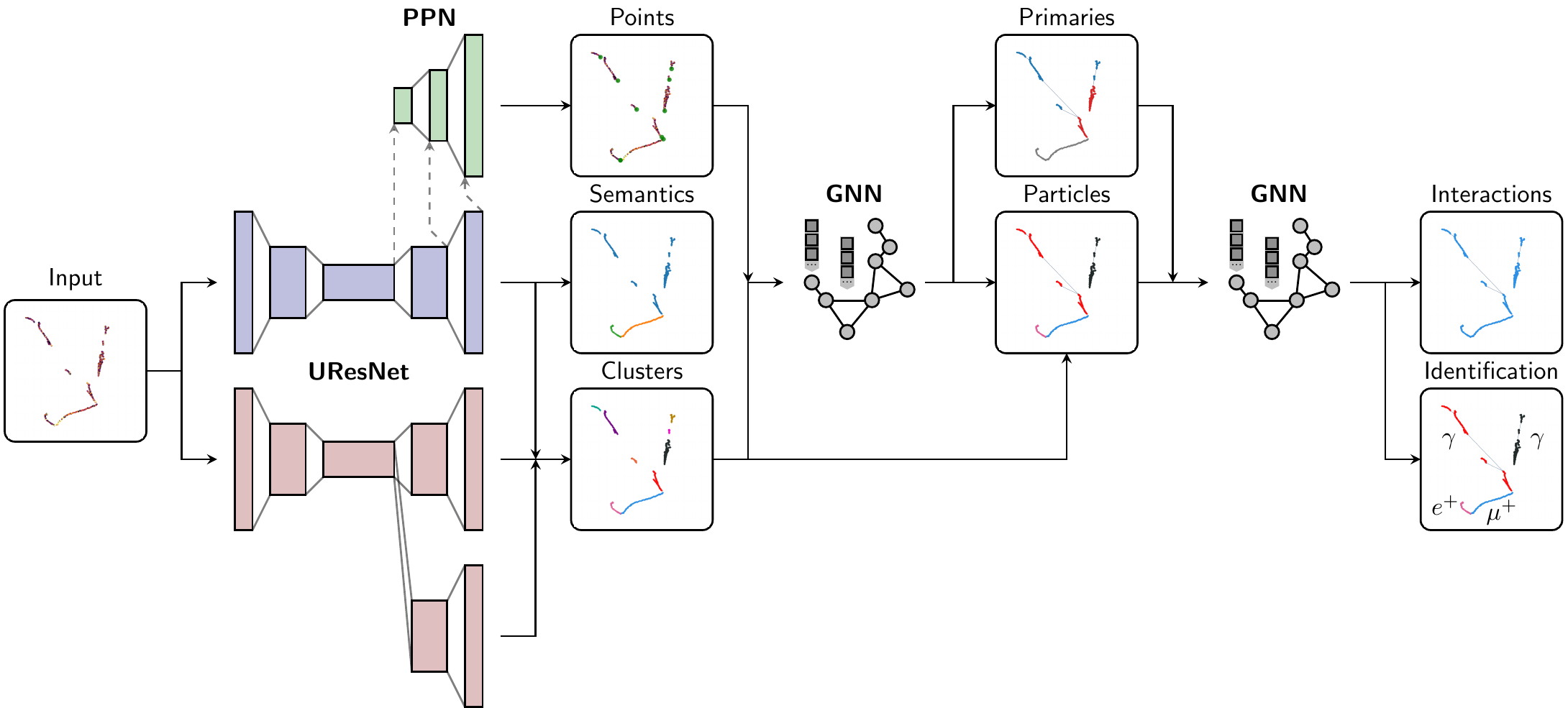}
    \caption{Schematic architecture of the end-to-end, ML-based reconstruction chain for LArTPCs.}
    \label{fig:chain}
\end{figure*}

Traditional pattern recognition techniques are heavily challenged by the extraordinary level of detail LArTPCs images exhibit. Machine Learning (ML) is at the forefront of computer vision and offers a way to cover the large phase space necessary for this task. This paper presents a novel, end-to-end, ML-based reconstruction chain for LArTPCs, schematically represented in figure~\ref{fig:chain}. The 3D input image is first passed through two parallel autoencoder networks designed to extract voxel-level features out of the image. The first network (blue), in conjunction with additional convolutions (green), classifies voxels in different abstract particle classes and identifies points of interest. The second (red) uses two decoders to build individual dense particle clusters within each aforementioned class. Two sequential graph networks  are then used to first assemble shower objects and identify primary fragments before aggregating particles into interactions and identifying their species.

The following sections detail the architectures of the network modules. Each stage of the reconstruction is trained and tested on the PILArNet dataset of 125280/22439 (train/test) rasterized 3D images of size $768^3$\,voxels capturing a realistic density of particle interactions in a 12\,m$^3$ volume of LAr~\cite{pilarnet}.

\section{Semantic Segmentation and Point Proposal}
The first module in the reconstruction chain is designed to identify the abstract particle type of each voxel~\cite{sparse-uresnet} and the location of important points~\cite{ppn}. These two tasks share a common backbone architecture called ``Sparse U-ResNet'':  a U-Net~\cite{unet} -- composed of a down-sampling encoder and an up-sampling decoder extracting features at various scales, i.e. \emph{depth} -- where convolutions have been substituted for ResNet blocks~\cite{resnet} implemented in the sparse convolutional network (SCN) framework~\cite{scn}. SCN makes deep convolutional neural network scalable to large 3D images -- including those encompassing the entire volume of a 10\,kton LArTPC used in the DUNE far detector~\cite{dune} -- as the computational complexity of sparse convolutions only increases with the number of active voxels. For the segmentation task, the output layer predicts a score for each of the target particle classes: electromagnetic shower, track-like, Michel electron, delta ray or low energy (LE).

Parallel to the U-shaped network, additional convolution layers are introduced at three spatial resolutions to form the so-called Point Proposal Network (PPN)~\cite{ppn}. Inspired by Region Proposal Networks~\cite{faster-rcnn}, the first two PPN layers attempt to predict a positive score for voxels that contain a ground-truth PPN point. Positive voxels form a mask that is then applied to the following PPN layers. For each voxel that has been selected through these successive attention masks, the final layer uses $3\times3$ convolutions to predict the point positions relative to the voxel centers and their particle classes.

\begin{figure}[htb!]
    \centering
    \hfill{}
    \subfloat{\includegraphics[width=0.508\linewidth]{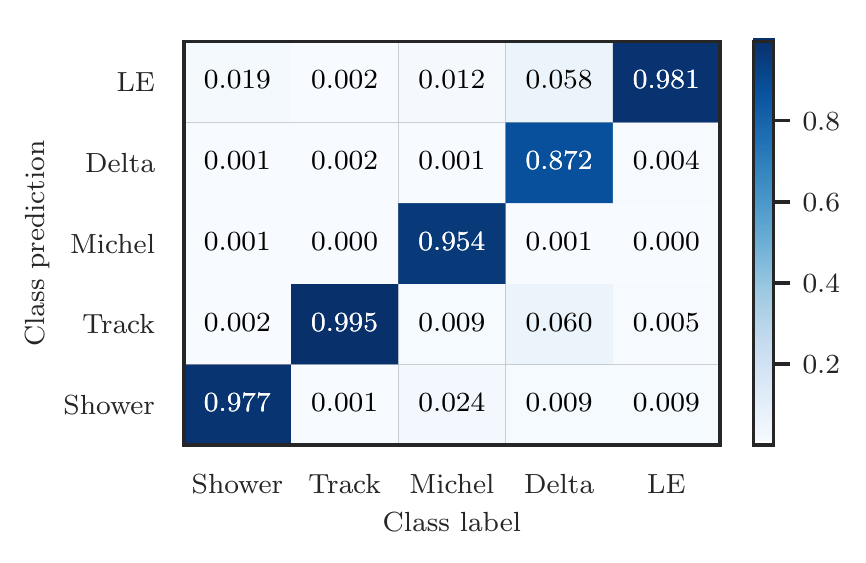}}
    \hfill{}
    \subfloat{\includegraphics[width=0.472\linewidth]{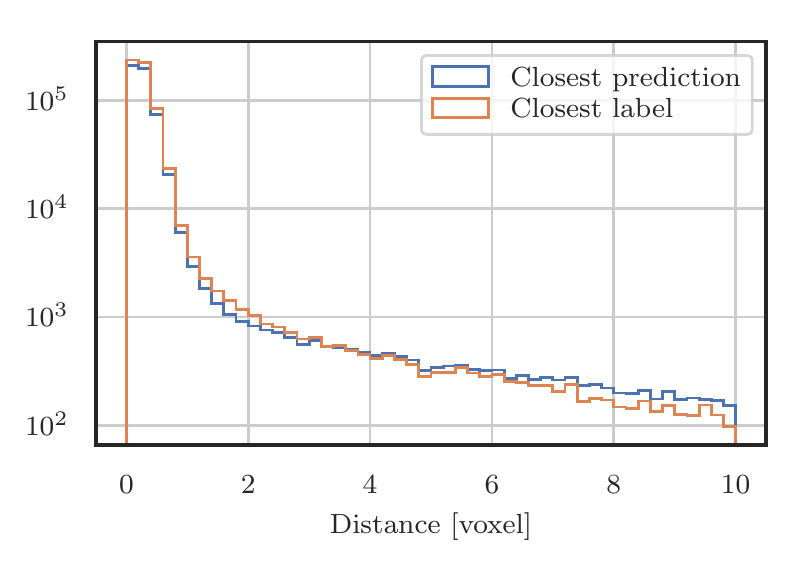}}
    \hfill{}
    \caption{(\textbf{Left}) UResNet semantic segmentation confusion matrix. Each column sums to 1. (\textbf{Right}) Distance from a ground-truth point to the closest predicted point (blue) and from a predicted point to the closest ground truth point (orange). Points of type delta are excluded from both histograms.}
    \label{fig:uresnet_ppn_metrics}
\end{figure}

The left panel of figure~\ref{fig:uresnet_ppn_metrics} shows the semantic segmentation confusion matrix. All the classes are identified with a high level of precision, with tracks and showers being classified with a voxel-wise accuracy of 97.7\,\% and 99.5\,\%, respectively. This algorithm shows a similar performance to previous results applying UResNet to 2D LArTPC images~\cite{uburesnet}. The largest source of confusion originates from delta rays misidentified as either track points or low energy depositions. The former mistakes can be explained by the overlapping nature of tracks and delta rays while the latter stems from labelling ambiguities that will be addressed in future datasets.

Point proposals are reconstructed by applying the point aggregation procedure described in \cite{ppn}. The right panel of figure~\ref{fig:uresnet_ppn_metrics} shows the distributions of distance from a true label point to the closest predicted point and vice versa. Traditional methods report 68~\% of neutrino interaction vertex reconstructed within 0.73~cm~\cite{pandora,ubvtx}. On the related task reported here, the PPN locates 68~\% of all points within a radius of 0.10~cm, and 95.9~\% of all points are found within 0.7~cm.  

\section{Dense Clustering}

The next module in the reconstruction chain involves clustering densely connected voxels into different particle instances~\cite{spatial_embeddings}. Only those voxels that belong to a common semantic class, as predicted by the semantic segmentation task, may be clustered together. This proposal-free method uses another U-ResNet with two decoders. The input image is passed through a shared encoder and then expanded into two output feature planes. The first feature plane is referred to as the \emph{embedding} layer and the second as the \emph{seediness} layer. The embedding layer learns a coordinate transformation of the input image voxel coordinates such that points that belong to the same cluster, i.e. particle instance, are embedded close to one another. The seediness layer quantifies how likely a given voxel is to be close to the centroid of embedded points that share the same cluster id. Once a reasonable transformation that groups voxels into localized clusters in embedding space is obtained, information from the two feature maps are combined to assign cluster labels in a sequential manner.

\begin{figure}[h]
    \centering
    \includegraphics[width=.9\textwidth]{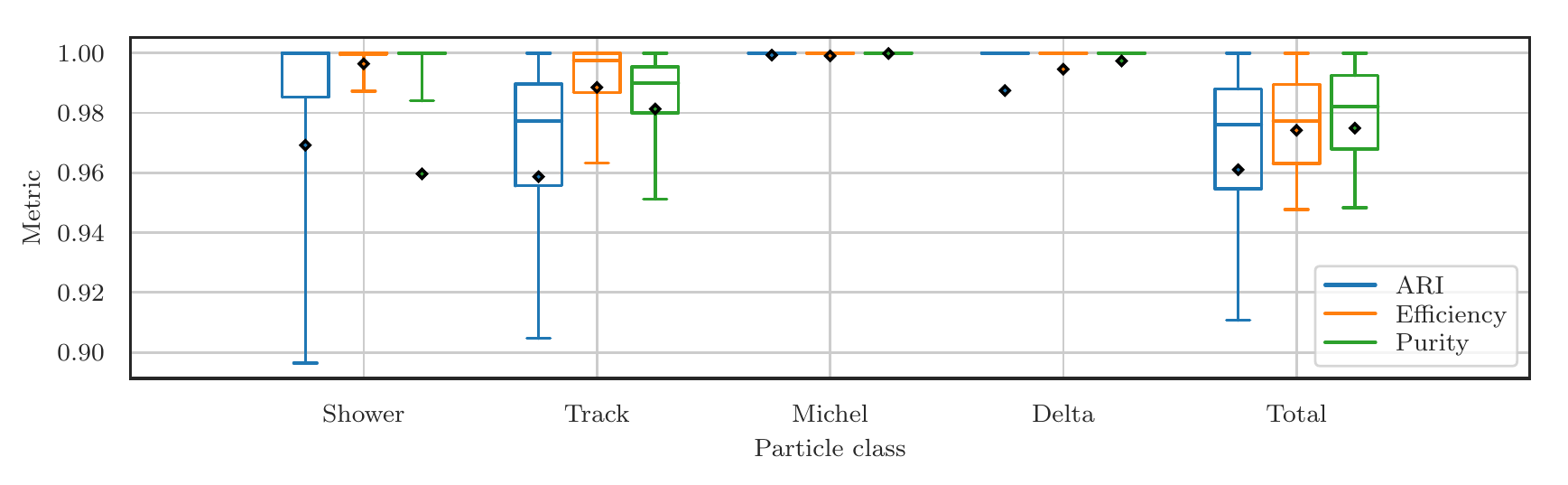}
    \caption{Box plot of the clustering metrics obtained with the UResNet-based dense particle clustering network for each class and all classes combined. The diamonds represent the means, the lines the medians, the boxes the IQRs and the whiskers extend from the $10^\text{th}$ to the $90^\text{th}$ percentile.}
    \label{fig:cluster_box}
\end{figure}

Three metrics described in~\cite{spatial_embeddings} are used to characterize the clustering performance: efficiency, purity and Adjusted Rand Index (ARI). Figure~\ref{fig:cluster_box} shows summary statistics of the clustering metric distributions associated with each semantic class. This algorithm achieves an average efficiency and purity of 97.5\,\% and a mean ARI of 96.1\,\% for all classes combined. At this stage, showers are clustered as locally dense fragments to be assembled by an algorithm described in the next section.

\section{Particle Aggregation}

At this stage of the reconstruction, Graph Neural Networks (GNNs) are used to cluster distant objects into superstructures: shower fragments into shower instances and particles into interactions~\cite{gnn_paper}. The particle instances are initially encoded into nodes each characterized by geometrical summary statistics, including the most likely PPN point in an instance. Nodes are connected together by a complete graph in which edges are each provided with geometric features related to the displacement vector separating the fragments it connects.

The node and edge features are updated through multiple message passing steps~\cite{inductive_bias}, after which the final edge and node features are reduced to a {\em node} score and an {\em adjacency} score matrix for edges. The ground-truth adjacency matrix is built such that if two nodes belong to the same group, the edge that connects them is given a label of 1. The edge scores are used to constrain the connectivity graph of particles, and the node scores to identify shower primaries or particle species. The edge selection mechanism, described in~\cite{gnn_paper}, optimizes a global graph partition score at the inference stage, which significantly increase clustering accuracy, compared a naive edge-wise classification.

\begin{figure}[!htb]
  \centering
  \hfill{}
  \subfloat{\includegraphics[width=0.49\linewidth]{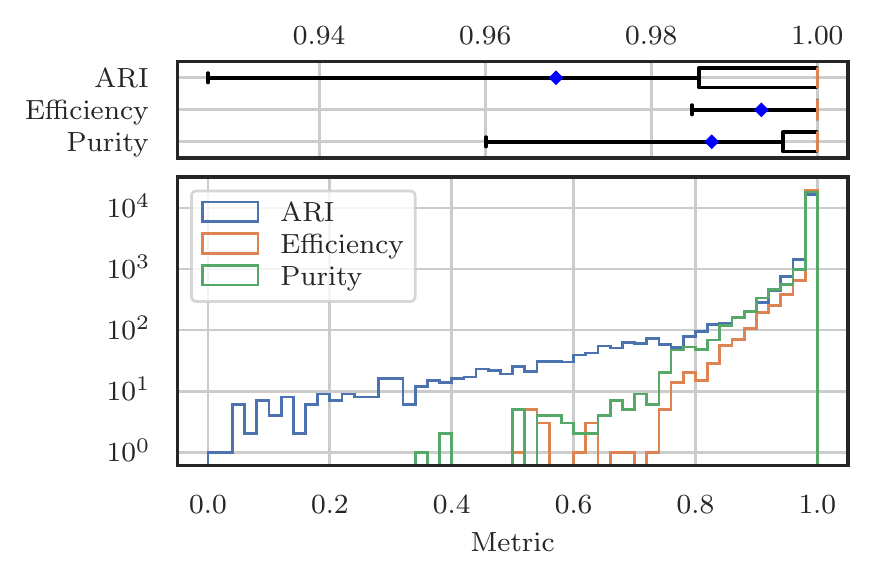}}
  \hfill{}
  \subfloat{\includegraphics[width=0.49\linewidth]{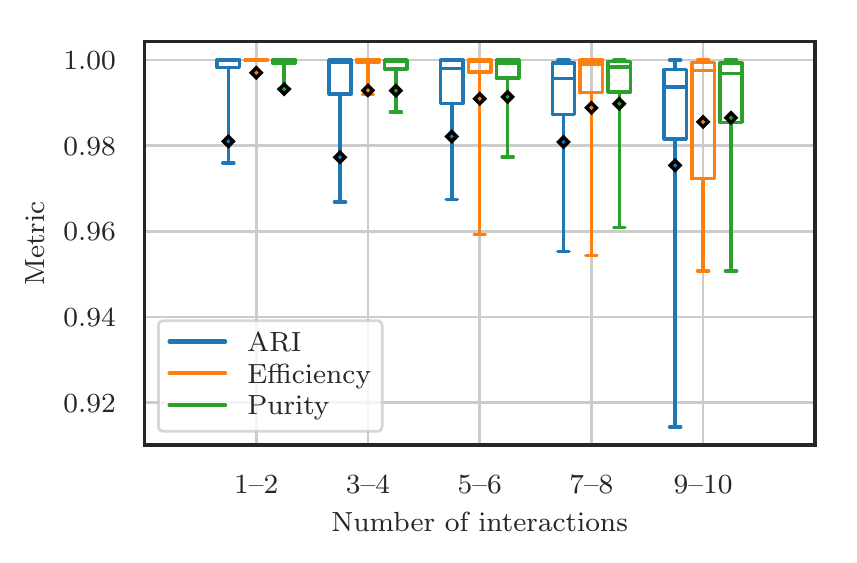}}
  \hfill{}
  \caption{(\textbf{Left}) Distribution of shower clustering metrics. (\textbf{Right}) Box plot of interaction clustering metrics as a function of the number of interactions per image. The diamonds represent the means, the lines the medians, the boxes the IQRs and the whiskers extend from the $10^\text{th}$ to the $90^\text{th}$ percentile.}
  \label{fig:gnn_clustering}
\end{figure}

The left panel of figure~\ref{fig:gnn_clustering} shows the distribution of shower clustering metrics on the entire test set. The network achieves an average purity of 98.7\,\%, an efficiency of 99.3\,\% and a mean ARI of 96.9\,\%. The network identifies primaries by selecting the node with the highest primary score in each predicted group, which yields a 99.5\,\% primary prediction accuracy for this dataset.

The right panel of figure~\ref{fig:gnn_clustering} shows summary statistics of the interaction clustering metrics on the entire test set for different number of interactions in the image. For the particle density of the DUNE near detector (DUNE-ND), the network achieves an average purity and efficiency of 99.1\,\% and a mean ARI of 98.2\,\%. This algorithm thus addresses one of the main challenges of the DUNE-ND, which will experience an unprecedented pile-up of neutrino interactions in a LArTPC.

\section{Conclusions}
This paper demonstrates the success of a modular, end-to-end, ML-based reconstruction chain which takes 3D particle interaction images as an input and hierarchically extracts increasingly high-level information at each stage by building upon the previous steps. The algorithm provides a comprehensive description of an event: a list of interactions per image, a list of particle instances per interaction, individual particle types and trajectories, start and end points. The performance of each module has been evaluated independently and shown to perform up to the highest standards. This reconstruction chain is the first crucial step towards a neutrino oscillation inference machine, which will employ a completely differentiable simulation pipeline upstream to infer oscillation parameters.

The computing resources needed for this chain scale with the number of non-zero voxels and not the image dimensions, making it the ideal candidate for LArTPC image data which are extremely sparse -- with over 99~\% of inactive voxels -- but contain locally-dense particle trajectories. The chain is also automatically end-to-end optimizable within a week using a single NVIDIA V100 GPU, or even faster when leveraging a distributed system. In comparison, the traditional approach involves months to years of manual software adjustments per data production campaign, which are sometimes run yearly in neutrino experiments. The chain presented in this paper is to be employed in future cutting-edge neutrino endeavors, including the SBN program and the DUNE experiment.

\section*{Broader Impact}

While ML methods could have a major impact in the way data analysis is handled in the scientific community, it is not expected to have any ethical or societal consequences. Nobody outside of the field of fundamental physics is expected to be affected by a technological upgrade in LArTPC event processing. The consequences of failure of this reconstruction chain are identical to that of one employing a traditional programming approach: inaccurate or biased scientific results, incorrect conclusions drawn from said results and, as a consequence, a misinformed community. However, the approach described in this document mitigates the main shortfalls of simple image classifiers and regression neural networks being used in physics today, which both suffer from large single-steps of information reduction. The chain presented here is composed of a series of task-specific neural networks which hierarchically extract increasingly high-level information out of the data, which allows one to both identify the short comings of the algorithm, and also dramatically reduces the impact of the training set. It is believed that this work will have an unequivocally positive effect.

\begin{ack}
This work is carried out as part of the HEP Advanced Tracking Algorithms at the Exascale (Exa.TrkX) project, and is supported by the U.S. Department of Energy, Office of Science, Office of High Energy Physics, and the Early Career Research Program under Contract DEAC02-76SF00515.
\end{ack}


\end{document}